\documentclass[10pt]{article}
\usepackage{graphicx}
\newcommand{\eq}{\begin{equation}}
\newcommand{\feq}{\end{equation}}
\newcommand{\eqn}{\begin{eqnarray}}
\newcommand{\feqn}{\end{eqnarray}}
\newcommand{\arr}{\begin{eqnarray*}}
\newcommand{\farr}{\end{eqnarray*}}
\newcommand{\beq}{\begin{equation}}
\newcommand{\eeq}{\end{equation}}
\newcommand{\bea}{\begin{eqnarray}}
\newcommand{\eea}{\end{eqnarray}}
\newcommand{\lb}{\label}

\def\*{{\bf ***}}

\def\phi{\varphi}

\def\vphi{\varphi}

\def\o+{\oplus}

\def\<{\langle}
\def\>{\rangle}

\def\interno{\hskip 2pt \vbox{\hbox{\vbox to .18
truecm{\vfill\hbox to .25 truecm
{\hfill\hfill}\vfill}\vrule}\hrule}\hskip 2 pt}

\def\({\left(}
\def\){\right)}
\def\[{\left[}
\def\]{\right]}
\def\=#1{\bar #1}
\def\~#1{\widetilde #1}
\def\.#1{\dot #1}
\def\^#1{\widehat #1}

\def\vphi{\varphi}
\begin{document}

\title{A symmetry breaking mechanism for selecting  the speed of
relativistic solitons\thanks{Work
supported in part by the Italian MIUR under the program COFIN2004,
as part of the PRIN project {\it ``Mathematical Models for DNA
Dynamics ($M^2 \times D^2$)''}.}}

\author{Mariano Cadoni ¥\footnote{email: 
mariano.cadoni@ca.infn.it},\, 
Roberto De  Leo  \footnote{email:roberto.deleo@ca.infn.it},\\ 
{\small Dipartimento di Fisica, Universit\`a di Cagliari  and  
INFN, Sezione di Cagliari} \\
{\small Cittadella Universitaria, 09042 Monserrato, Italy}\\
Giuseppe Gaeta \footnote{email:gaeta@mat.unimi.it}\\ 
{\small Dipartimento di
Matematica, Universit\`a di Milano, }\\{\small via Saldini 50, 20133
Milano, Italy}}

\vfill

\maketitle

{\bf Abstract.}
We  propose a mechanism for fixing the velocity of
relativistic soliton based on the breaking of the Lorentz symmetry
of the sine-Gordon (SG)  model. The proposal is first elaborated for a
molecular chain model, as the simple pendulum limit of a double 
pendulums chain. It  is then 
generalized to a full class of two-dimensional field theories of the sine-Gordon
type.  From a phenomenological point of view, the mechanism allows one to
select the speed of a SG soliton just by tuning  elastic couplings
constants and  kinematical parameters.
From  a fundamental,
field-theoretical point of view we show
that the  characterizing features of  relativistic SG solitons 
(existence of conserved topological charges  and stability) may be still
preserved even if the Lorentz symmetry is broken and a soliton of a
given speed is selected.

\section{Introduction}
In the past decades solitons have become a very useful concept in
many areas of mathematical, theoretical and applied physics.
Originally discovered as  solitary waves
solutions of non linear partial differential equations \cite{russell,kdv,zk},
solitons have
become a generic paradigm of non linear physical phenomena 
\cite{daupey,dodd}.

Of particular relevance, both from the theoretical and
phenomenological point of view, are relativistic solitons, in
particular topological two-dimensional (2D) sine-Gordon (SG)
solitons (kinks)\cite{rubinstein}, on which this paper will be
mainly focused. On the phenomenological side, SG solitons have
wide range of applicability:  non linear molecular \cite{Dav} and
DNA dynamics \cite{Eng,YakuBook,GRPD,PeyNLN}, Josephson effect
\cite{barone}, ferromagnetic waves \cite{feld, bar}, non linear optics
\cite{lamb,taylor}, superconductivity \cite{cattuto} and many
others \cite{manton}.

From a theoretical point of view SG solitons exhibit  remarkable
analogies with relativistic particles. They share with the latter
the feature of being localized, non dispersive objects realizing
the Lorentz symmetry. SG solitons are characterized by short range
forces and they can interact without loosing their identities
\cite{lamb1}.

Moreover, topological SG solitons are stable and their existence
is related to the discrete internal symmetries (and  the existence
of degenerate vacua) of the two-dimensional (2D) field theory of
which they are classical solutions. Topology offers an elegant way
of classifying solitons in terms of the homotopy group of the
mappings between the vacua and the asymptotical  field
configurations \cite{rubinstein}.

The soliton-particle analogy has been very fruitful both for
particle physics and quantum field theory. Starting with the
pioneering  work of Finkelstein and Misner \cite{fm}, Rubinstein
\cite{rubinstein} and Skyrme \cite{skyrme}, the analogy has found
applications in various contexts such as S-Matrix formalism
\cite{zamo}, current algebra effective Lagrangians \cite{witten},
supersymmetric quantum mechanics \cite{witten1}, black hole
physics \cite{Gegenberg:1997ns,Cadoni:1998ej} just to mention a
few of them.

Solitons have also became central in the study of integrable PDEs
\cite{calogero,Zakh} and in many geometrical matters (in
particular connected with Field Theory), see e.g.
\cite{DNF}.

A  general feature of relativistic solitons and in particular of
SG solitons, is the fact that the soliton speed $v$ is a free
parameter, which can be fixed  by choosing  initial conditions
and is bounded from above by a limiting value $c_{0}$. This is
obvious in view of the Lorentz symmetry of the model: only the
limiting upper bound (corresponding to the speed of light for a
relativistic particle) of the soliton speed is fixed by the model,
whereas $v$ can be changed applying  a boost.

This degeneracy in $v$ is a necessary  property of a
relativistic field theory such as SG theory and appears rather
natural in view of the soliton-particle analogy. However, in 
phenomenological situations where the experiments give a well-defined 
value for the propagation speed of the soliton
it represents a loss of predictive
power of the model. By modelling some non-linear dynamical process
using a SG-like field theory very often it is crucial not only
that solitary, non dispersive waves do exist but also that their
velocity is at least of the same order of magnitude of that
observed in experiments.

Thus, a selection mechanism for the soliton velocity should be
very welcome from the phenomenological point of view. In this
context an instructive example is represented by the use of
solitons in the description of DNA non linear torsion dynamics. It
is  believed that kink solitons of the SG type play a
crucial role in the DNA transcription process \cite{Eng,
YakuBook}. In the past decades several models of DNA torsional
dynamics have been proposed that allow for the existence of SG
kink solitons ( see e.g \cite{YakuBook,GRPD} and references therein). 
However, in most of
them the soliton velocity is not fixed by the model\footnote{The
only exception is the model of Ref. \cite{CDG}, where the
selection mechanism discussed in this paper was first observed but 
not identified in its generality.},
whereas the DNA transcription process, which the soliton
propagation is assumed to model, occurs with a speed of a
well-defined order of magnitude.

Although desirable from a phenomenological point of view, it is
not very easy  to implement a general selection mechanism for the
soliton speed. The high sensibility of non linear systems to the
initial conditions,  practically rules out  a selection mechanism
based entirely on  tuning  initial conditions of the system.

On the other hand, one expects a speed selection mechanism to have
a strong impact on the existence and on  fundamental properties of 
SG solitons. In
fact any velocity selection  breaks the Lorentz symmetry of the
model. It is not {\it a priori} evident that after
this breaking  the theory will still allow for localized solutions 
with the 
features  of the SG solitons such as stability and 
topological classification.

In this paper we will present and discuss in detail a  velocity
selection mechanism for SG topological solitons.
Our starting point is  a result we found in previous investigation
of a composite model for DNA torsional dynamics, which is essentially a
double chain of coupled pendulums \cite{CDG,CDG1}. We found that in the
limiting case  where one of the  two  coordinates describing the angular
displacement of the pendulums is constrained to be identically zero, 
the continuum
limit of the model has SG solitonic solutions with speed fixed in
terms of the elastic and inertial parameters of the chain.

The possibility of selecting the speed of SG solitons by adding to 
the Lagrangian terms breaking the Lorentz symmetry has been first 
recognized in Ref. \cite{mpey}. A speed selection mechanism has been 
also found in the case of solitonic solutions of the non-linear 
Schr\"odinger equation \cite{harikumar}. In particular, in Ref. 
\cite{mpey} the authors consider a SG model modified by the addition 
of  further  kinetic and potential terms. In this paper we consider 
instead a two fields generalization of the  usual SG model. Moreover 
the breaking of the Lorentz symmetry, which is responsible for the 
fixing of the soliton speed is rather subtle. It is generated 
by the presence of two 
(instead of one)  upper 
limit speeds for the propagation of waves.      

The {\bf plan of the paper} is as follows. We will first review 
the results of Ref.  \cite{CDG}  and  reformulate them 
as  the simple pendulum limit of a double pendulums molecular chain model 
(Section \ref{S2}).
 We will show
that the selection mechanism is related to the existence of a
conditionally conserved quantity, a conservation law that holds only
when the system is constrained in a limited region of its phase
space, i.e. to a region of the reduced one-dimensional (1D) equations of
motion describing propagating waves (Section \ref{S3}).  We will then
pass to consider the most general model for which the selection
speed mechanism does work. We will show that the general model can
be formulated as a two fields generalization of 2D SG field theory
(Section \ref{S4}). We will consider a simplified, minimal version of
the model, which has the nice feature of being explicitly solvable
in its general form (Section \ref{S5}). The breaking of the Lorentz
symmetry is discussed and it is shown that it is related to the
presence of two (instead of one) upper bound limiting speeds
(Section \ref{S6}). The discrete symmetries of the model are  discussed
and used for the topological classification of the kinks (Section
\ref{S7}).  The stability of our solitonic solutions are also
investigate.  In particular we show that the Lorentz symmetry
breaking term acts as damping term for linear perturbation near
the soliton (Section \ref{S8}).  Finally we state our conclusions
(Section \ref{S9}).

\section{Soliton  speed  selection   in  molecular chain \\ models}
\lb{S2}

Let us start from  the double pendulums chain model of Ref.
\cite{CDG} modeling DNA torsional dynamics. The two DNA helices
are modeled by two chains of double pendulums. At each, equally
spaced with distance $\delta$, site of the two chains there is a
first pendulum, which is a disk radius $R$ and momentum of inertia
$I$. Attached to the border of the disk there is second pendulum
of length $r$ and mass $m$. The rotation angles of the two
pendulums with respect to the  equilibrium positions are denoted
respectively by $\theta^{a}_{i}$, $\vphi^{a}_{i}$, where the index
$a=1,2$ refers to elements in the two chains and $i\in {\bf Z}$
identifies the site in the chain and $\theta^{a}_{i},
\vphi^{a}_{i}\in S^1$. The pendulums
interact with nearest-neighbor interactions of three types:
stacking (coupling between successive pendulums on the same
chain), pairing (coupling between pendulums on the same site but
opposite chain) and torsional (coupling between successive disks
on the same chain), characterized by coupling (elastic) constants
given respectively by $K_{s},K_{p},K_{t}$. Moreover, the second 
pendulum is not free to swing through a full circle but it is 
instead constrained to stay in a range $\vphi^{a}_{i}\le\vphi_{0}<2\pi$. 
This constraint can be modelled by adding a confining potential
$V_{c}(\vphi)$, which has the effect of limiting {\sl de facto} the 
excursion of the $\vphi$ angles;
see Ref. \cite{CDG,CDG1}
for details.

As here we are interested not in the details of the model but in
its general behavior, we will make, with respect to Ref.
\cite{CDG,CDG1}, a number of simplifying assumptions. First we
will consider from the beginning only symmetric solutions,
$\theta^{1}_{i}= \theta^{2}_{i}:= \theta_{i}$, $\vphi^{1}_{i}=
\vphi^{2}_{i}:=\vphi_{i}$. This position reduces the degrees of
freedom from four to two per site. Second, to reduce the number of
kinematical parameters we will set $R=r$. Under these simplifying 
assumptions, in the
continuum ($\delta\to 0$) limit the dynamics of the model is
described by a 2D field theory with Lagrangian density
\bea\lb{lagra1} {\cal{L}}&=&\frac{1}{2}\left\{
I\theta_{t}^{2}-\omega_{t}\theta_{x}^{2}+ r^{2}\left[ m
\theta_{t}^{2}-\omega_{s}\theta_{x}^{2}+ 2 \cos
(\vphi)\left(m\theta_{t}(\vphi_{t}+\theta_{t})\right. \right.\right.+
\nonumber\\
&-&\left. \left.\left.\omega_{s}\theta_{x}(\vphi_{x}+
\theta_{x})\right)
+m(\vphi_{t}+\theta_{t})^{2}-\omega_{s}(\vphi_{x}+\theta_{x})^{2}
\right]\right\}+\nonumber\\
&+&4r^{2} K_{p}\left( \cos\theta +\cos(\vphi+\theta)  -
\frac{1}{2}\cos(\vphi) -\frac{3}{2})\right)- V_{c}(\vphi), \eea
$x$ is the spatial coordinate along the chain, the fields
$\theta(t,x)$ and $\vphi(t,x)$ represent the angular coordinate
$\theta_{i}$ and $\vphi_{i}$ in the continuum limit, the indices
$x$ and $t$ denote partial derivation and
$\omega_{t}=K_{t}\delta^{2} ,\, \omega_{s}=K_{s}\delta^{2} $.
We will use in the Lagrangian (\ref{lagra1}) a rather generic 
confining potential $V_{c}$. We will only assume that $V_{c}(\vphi)$ 
is an even function of $\vphi$, which has a minimum at $\vphi=0$.
Moreover, we can take without loss of generality $V_{c}(0)=0$

At this point the interpretation of the Lagrangian (\ref{lagra1})
as modelling the DNA torsional dynamics is not anymore compulsory,
nor so relevant. We can as well consider the Lagrangian as
describing the continuum limit of the torsional dynamics of a
single molecular chain made of two pendulums.  The pairing
interaction for the DNA double chain becomes an external potential
for the pendulums: $V=-4r^{2} K_{p}\left( \cos\theta
+\cos(\vphi+\theta) - \frac{1}{2}\cos\vphi -3/2\right)$,
whereas the pairing and torsional interaction generate in the
Lagrangian (\ref{lagra1}) the gradient ($x$-derivative) terms.

Let us  consider travelling waves solutions of the field theory
(\ref{lagra1}), i.e solutions depending only on $z=x+vt$ with
fixed speed $v$: $\theta = \theta(z)$, $\vphi = \vphi(z)$. With
this position the field equations stemming from (\ref{lagra1})
become
\beq\lb{l}
\begin{array}{l}
\mu  \, \vphi'' \ + \ \mu  (1 +  \cos \vphi ) \, \theta'' \ = \\
\ \ \ \ = \ - 4  K_p  \, \sin \(\vphi + \theta \)  -\mu \, \sin \(
\vphi \) \, (\theta' )^2 \ + \ 2 K_p  \, \sin (\vphi) -\frac{\partial 
V_{c}}{\partial \vphi}
 \ ; \\
\mu  (1 +  \cos \vphi ) \, \vphi'' \ + \ [ (J /r^2) + 2\mu (1 +
 \cos \vphi )] \, \theta'' \ = \\
\ \ \ \ = \ - 4  K_p \(  \sin \theta + \sin (\vphi + \theta) \) +
\mu  \, \sin (\vphi) [ (\vphi')^2 + 2 \vphi' \theta' ] \ ,
\end{array} \feq
where the prime denotes derivatives with respect to $z$ and
following the notation of Ref. \cite{CDG}
we define
\beq\lb{6} \mu \ := \ (m v^2 - \omega_s ) \ , \ \ J \
:= (I v^2 - \omega_t  ) \ . \feq
In general the system (\ref{l}) cannot be solved analytically in
closed form. One has to resort to numerical calculations in order to
show that the system admits solitonic solutions \cite{CDG}.
However,  in Ref. \cite{CDG} it was shown that one can  find a
 solitonic solution with {\sl fixed speed}, just by freezing the angle
$\vphi$, i.e by setting $\vphi=0$.
We note that if we force $\vphi (z)= 0$,  we are actually 
considering a chain of simple pendulums,
i.e. a sine-Gordon equation. This constraint can be accommodated 
in our setting in a dynamical way, by acting on the confining potential
$V_c$: this should be made stronger and stronger 
and the maximum angle $\vphi_0$
smaller and smaller. In the limit $\vphi_0 \to 0^+$ and $(\partial 
^{2}¥V_{c}/\partial \vphi^{2}) (0)
\to + \infty$, we expect to recover the solitons of the
sine-Gordon equation.

Setting $\vphi=0$ and using $(\partial ¥V_{c}/\partial \vphi) (0)=0$, the system
(\ref{l}) is equivalent to the SG field equation
\beq\lb{ge1}
\mu\theta''=-2K_{p}\sin\theta,
\feq
with the compatibility condition between the two equations of the
system (\ref{l}) given by
\beq\lb{comp}
J=0.
\feq
In principle one could consider a  generic constraint $\phi= q 
\theta$, where $q$ is a proportionality constant. However, it is not 
difficult to realize that for $q\neq 0$ such a constraint, when used in 
Eqs. (\ref{l}),  will lead to  two different determinations of 
$\theta$ making the  two resulting equations incompatible.

Eq. (\ref{ge1}) has to be integrated with the usual boundary
conditions for kinks
$\theta(\pm\infty)=2n_{\pm}\pi,\,\theta'(\pm\infty)=0$ with
$n_{\pm}\in{\bf Z}$. Here $n=n_{+}-n_{-}$ is the kink winding
number. For $n=1$ Eq. (\ref{ge1}) can be easily integrated to give
the kink
\beq\lb{fe3} \theta_{0}= 4 \arctan[e^{ {\cal K}z}],\quad\quad
\vphi=0,\quad {\cal K}= \sqrt{\frac{2K_{p}}{|\mu|}}, \feq
whereas the compatibility condition (\ref{comp}) fixes the speed
of propagation of the soliton to the speed $c_{t}$ of the
transverse sound waves supported by the elastic torsional forces
acting on the disk \beq\lb{speed} v=c_{t}=\sqrt{\omega_{t}/I}.
\feq Moreover, because the soliton exist only for $\mu<0$, the
soliton speed is bounded from above by the speed $c_{s}$ of
transverse sound waves supported by the elastic stacking forces
acting on the pendulum, i.e. \beq\lb{speed1}
 v \le c_{s}=\sqrt{\omega_{s}/m}.
 \feq
In view of Eq. (\ref{speed}),
 this implies the following constraint on the stacking, torsional coupling
 constants and kinematical parameters of the system:
\beq\lb{h8}
\frac{K_{t}}{I}<\frac{K_{s}}{m}.
\feq

The condition $\vphi=0$ fixes the speeds of the soliton
to the value (\ref{speed}) also for kinks with  winding
numbers $n>1$. In fact,  in Sect.  \ref{S7} we will show that the fixed 
speed solitonic solutions of Eq. (\ref{ge1}) allow for the same  topological 
classification of the usual SG solitons.

The selection  mechanism for the soliton speed discussed
above represents  a very nice and simple  way to produce
SG solitons with a given speed in double pendulums molecular chains. To select the
soliton speed one just needs to tune the torsional and stacking
coupling constants
and the kinematical parameters of the chain such that Eqs.
(\ref{speed}) and (\ref{h8}) are satisfied. 
Acting on the confining potential
$V_c$, making it stronger and stronger,  one  obtains the 
single pendulum limit of the  
double pendulums chain.
The angle is frozen to  $\vphi=0$ and  a SG
soliton with a speed equal to that of the transverse sound waves supported by
the torsional forces acting on the disk is selected.
Notice that the above described mechanism can be obviously used to 
devise and realize non-linear media where solitons propagate  at a 
given fixed speed.

Although very simple the mechanism we are proposing raises a
number of questions.  How is our mechanism related to fundamental
features of the SG system such as integrability,  Lorentz symmetry
and topological classification? Can our mechanism be generalized
to a broad class of models or is it just an artifact of the
peculiar couplings present in the Lagrangian (\ref{lagra1})? Is
the SG soliton we are selecting stable? We will try to answer
these questions in the following sections.

\section{Integrability and conditionally  conserved \\ quantities}
\lb{S3}

The simple (one single degree of freedom)  SG system (\ref{ge1})
is integrable, in the sense that the initial value problem can 
always be solved analytically in closed form. Moreover, this
integrability is related to the existence of a conserved quantity
(a  first integral of the equations of motion), which can be
easily identified as the energy of the system.

On the other hand, the (two degrees of freedom) system (\ref{l})
in general allows for only one conserved quantity -- the energy --
and it is not integrable in the sense used above. Nonetheless, it
can be solved analytically in closed form and reduces to the SG
system (\ref{ge1}) if we limit ourselves to consider only
$\vphi=0$ solutions. One is therefore led to search for 
a conditionally conserved quantity (a discussion of this weak 
form of integrability can be found in Ref. \cite{sarlet}).
That is, a quantity that in general is not conserved by the
dynamical evolution of the system, but which becomes conserved
when the dynamics is restricted to the subspace $\vphi=0$.

To this aim, let us first note that the equations of motion
(\ref{l}) describe a dynamical system with two degrees of freedom,
with the coordinate $z$ playing the role of the time. The equation
of motion can be derived from the Lagrangian $L=T-V$,
\bea\lb{lagra} L&=& \frac{1}{2} J (\theta')^{2}+ \frac{\mu
r^{2}}{2}\left[ (\theta')^{2}+2 \cos(\vphi) \left(
(\theta')^{2}+\theta'\vphi'\right)+\left(\theta'+\vphi'\right)^{2} 
\right] \nonumber\\
&-& 2 K_p r^{2} \left[  \cos\vphi -2 \cos (\theta+\vphi)- 2
\cos\theta\right]-V_{c}(\vphi). \eea
With a little algebra we obtain from the system (\ref{l})  the
equation,
\beq\lb{ce} \frac{d}{dz} P(\theta',\vphi',\vphi)=
{\cal{F}}\left(\theta,\vphi,(\theta')^{2} ,(\vphi')^{2}\right),
\feq
where ${\cal{F}}$ is given by 
\beq\lb{F} {\cal{F}}=2r^{2}
\left[2K_{p}\left( \sin(\theta+\vphi)- \sin\theta-\sin
\vphi\right)+\mu \sin\vphi\left(
(\theta')^{2}+\vphi'\theta'\right)\right]+2 \frac{\partial V_{c}}{\partial 
\vphi}¥. \feq and $P$ is linear
in the momenta $P_{\theta}=\delta L/\delta\theta',\quad
P_{\vphi}=\delta L/\delta\vphi'$:

\beq\lb{P} P=P_{\theta}-2 P_{\vphi}= J \theta'+
s(\vphi)\mu\vphi',\quad s= r^{2} (\cos\vphi-1).
\feq
Notice that the coefficient $s$, its  derivative with respect to
$\vphi$  and ${\cal{F}}$ vanish when evaluated on $\vphi=0$. The
physical interpretation of Eq. (\ref{ce}) is very simple: this is
just Newton's second law with $P$ playing the role of the impulse
and ${\cal{F}}$ that of the force.

The equations of motion (\ref{l}) can be now rewritten in the
equivalent form
\bea\lb{l1}
\mu \vphi'' + \mu (1+ \cos \vphi ) \theta'' \ &=& \ - 4 K_p  \sin
\(\vphi + \theta \) - \mu
 \sin \( \vphi \)  (\theta' )^2 +
2 K_p  \sin (\vphi) \ \nonumber \\
P' \ &=& \ {\cal{F}}.
\eea
In general the momentum $P$ is not conserved, but becomes
conserved when ${\cal{F}}=0$. This can be achieved if we restrict
ourself to consider only the solutions of Eqs. (\ref{l1}) that
satisfy $\vphi=0$. Taking into account that ${\cal F}(0)=s(0)=s'(0)=0$ Eq.
(\ref{ce}) gives a  conservation law for
\beq\lb{cons} P(\vphi=0):= H=J\theta'. \feq
Thus, $H$ is the  conditionally conserved quantity we have been
searching for.

It can be interpreted has an ``effective'' angular momentum  of
the disk. In fact $J$ has two contributions, see Eqs. (\ref 6).
The first comes from the geometrical momentum of inertia of the
disk, the second has opposite sign and  comes from $K_{t}$, the
elastic forces acting on the disk.

For $\vphi=0$, Eqs. (\ref{l1}) reduce to
\beq\lb{fe1}
\mu\theta''=-2K_{p}\sin\theta,\quad\quad   \frac{d H}{dz}=0.
\feq
For
$J\neq 0$ the previous equations do not admit solitonic solutions
(actually the only solution is the trivial one $\theta=0$).
Solitonic solutions exist only for $J=0$, when the conservation
law is trivialized. In this case Eqs. (\ref{fe1}) coincide
with Eq.
(\ref{ge1}), (\ref{comp}), which give the fixed speed soliton
solution (\ref{fe3}).

Thus the physical mechanism behind the fixing of the soliton speed
is the following. When restricted to the subspace of solutions
$\vphi=0$ the system has  a conservation law, which is not
compatible with the existence of solitonic excitations. The
requirement of existence of solitons implies a trivialization of
the conservation equation, which in turn fixes the soliton speed.
From this point of view the condition $J=0$ can be seen as a sort
of resonance condition: both the force and the ``effective'' mass
in equation (\ref{ce}) must be zero.

\section{The general model as a 2D SG-like field theory}
\lb{S4}
The mechanism for selecting the soliton speed described in the
previous sections for the molecular chain model (\ref{lagra1}) is
rather generic. As we have shown in the previous section, the
speed selecting mechanism is related to the existence of a
conditionally  conserved  quantity and it is rather independent from
the specific form of the interactions characterizing the model.
Therefore, it will be present if the system satisfies some general
conditions.

\subsection{The general speed selection mechanism  in molecular chains}

Below is a list of the main ingredients that are needed for the
mechanism to work in the case of a  molecular chain. We need a
chain (array) of identical  mechanical systems with the following
features:
\begin{enumerate}
\item The system must have at least two degrees of freedom ($X,Y$)
characterized  by two masses (or momenta of inertia)  $m,M$ with $m\neq
M$;  
\item There are at least  two types of interactions: $a)$ An
elastic force (coupling constant $K_{t}$) originated by the
interaction between neighboring sites on  the chain; b) A non
linear external force (coupling constant $K_{p}$) acting on the
single site; 
\item
There is a confining potential $V_{c}$ that limits the range of 
variation of one degree of freedom (e.g $Y$)  
and allows  to freeze $Y$. That is by making
$V_{c}$ stronger and stronger we perform the dynamical reduction 
from two to one single degree of freedom  $X$;
\item 
Freezing the  degree of freedom  $Y$ we
have both a conservation law for the momenta conjugate to $X$
and solitonic solutions for $X$.
\end{enumerate}

If the  conditions above are satisfied, then the mechanism
described in the previous sections  will work. In particular,
considering the continuum limit of the chain and travelling wave
solutions $X(z)=X(x+vt),\, Y(z)=Y(x+vt)$  we get a system of two
differential equations which describe a mechanical system with two
degrees of freedom with effective masses \footnote{Depending on the 
particular model one is considering, the equations below may also 
assume a more general form $m_{{\rm eff}}=
v^{2}m - f(r_{i})K_{t}\delta^{2}$ (and similarly for $ M_{{\rm 
eff}}$),
where $f(r_{i}¥)$ is a function of the geometrical parameters $r_{i}$ 
characterizing the model.} 
\beq m_{{\rm eff}}=
v^{2}m - K_{t}\delta^{2},\quad M_{{\rm eff}}= v^{2}M -
K_{t}\delta^{2}. 
\feq
As the conservation of the momentum
conjugate to $X$ (for $Y$ frozen), $P_{X} =M_{{\rm eff}} dX/dz$ is
not compatible with the existence of solitons, this will require
$M_{{\rm eff}}=0$, which will fix the soliton speed in terms of
transverse phonon speed of the chain, 
\beq v=\delta\sqrt{\frac{K_{t}}{M}}.
\feq 
Moreover, $m_{{\rm eff}}$  is the analogous of the parameter $\mu$ in 
Eqs. (\ref{6}) and  (\ref{ge1}), hence the soliton will exist only for
$m_{{\rm eff}}<0$, 
implying $v<\delta\sqrt{K_{t}/m}$.
Consistency of these equations requires $M>m$.

\subsection{Speed selection in field theory}

The explicit  construction of models satisfying the previous
conditions becomes much simpler if we consider them directly at the
field theoretical, Lagrangian level, as generalization of the SG system, rather
then as originated from a molecular chain.
The main advantage of this approach is that the  speed selection mechanism
becomes a general, field theoretical, effect not necessarily
originated from the continuum limit of some discrete model.
In the rest of this paper we  will follow this approach and we will
make no more reference to molecular chain models.

To write down the Lagrangian for a general class of field theories
that exhibits the speed selection effect, we first redefine the
field $\vphi$ in the Lagrangian density (\ref{lagra1}). Defining
$\vphi=\Phi-\theta$ the Lagrangian (\ref{lagra1}) becomes
\bea\lb{sg} 
{\cal{L}}=\frac{1}{2}\left\{
I\theta_{t}^{2}-\omega_{t}\theta_{x}^{2}+
r^{2}\left[ m \theta_{t}^{2} -\omega_{s}\theta_{x}^{2}+2 \cos
(\Phi-\theta)\left(m\theta_{t}\Phi_{t}
-\omega_{s}\theta_{x}\Phi_{x}\right)+\right. \right.
\nonumber\\
\left. \left.
+m\Phi_{t}^{2}-\omega_{s}\Phi_{x}^{2}+ 8
K_{p}\left( \cos\theta +\cos\Phi  - \frac{1}{2}\cos(\Phi-\theta)
-\frac{3}{2}\right)\right]\right\}-V_{c}(\Phi-\theta). \eea
In this new  parametrization,  the fixed speed soliton solution
is given by $\Phi=\theta=\theta_{0}$, with $\theta_{0}$ given by Eq.
(\ref{fe3}).

Solitons are localized excitations of finite energy. The energy of
the solution (\ref{fe3}) can be calculated from the Hamiltonian
density. From the Lagrangian (\ref{sg}) one can easily obtain the
Hamiltonian density 
\bea\lb{ha} {\cal{H}}=\frac{1}{2}\left\{
I\theta_{t}^{2}+\omega_{t}\theta_{x}^{2}+ r^{2}\left[ m
\theta_{t}^{2}+\omega_{s}\theta_{x}^{2}+ 2 \cos
(\Phi-\theta)\left(m\theta_{t}\Phi_{t}
+\omega_{s}\theta_{x}\Phi_{x}\right)\right. \right.+
\nonumber\\
+\left. \left.
m\Phi_{t}^{2}+\omega_{s}\Phi_{x}^{2}- 8
K_{p}\left( \cos\theta +\cos\Phi - \frac{1}{2}\cos(\Phi-\theta) -
\frac{3}{2}\right)\right]\right\}+V_{c}(\Phi-\theta).\eea 
The constant field configuration of
minimal energy (vacua) can be obtained by minimizing the potential
term in Eq. (\ref{ha}). They   are given by $\theta= \Phi=2n\pi,\,
n\in {\bf Z}$ and  have  zero energy. The energy of the soliton
(\ref{fe3}), connecting vacua with $n_+ = n_- + 1$, is

\beq\lb{en} H=\int_{-\infty}^{\infty}  {\cal{H}} dx = 16 \, r^{2}
\, K_{p} \
\frac{\omega_{t}+8r^{2}\omega_{s}}{\omega_{s}-(m/I)\omega_{t}}.
\feq As expected the energy of the soliton  (measured with respect
to that of the vacuum) is finite and positive in the  range of
existence of the soliton.

It is useful, in particular in view of the discussion about the
Lorentz symmetry of the theory, to write  the Lagrangian using a
Minkowski spacetime notation. Introducing a spacetime  metric of
signature $(1,-1)$, the two fundamental speeds $c_{t},\,\,c_{s}$
(we assume $c_{t}<\,c_{s}$) given by Eqs (\ref{speed1}) and
respectively (\ref{speed}), and the derivative operators
\beq\lb{fe11} \hat\partial_{\nu}=\left(\frac{
\partial}{c_{t}\partial t}, \frac{ \partial}{\partial x}\right),
\quad
 \partial_{\nu}=\left(\frac{ \partial}{c_{s}\partial t},
\frac{ \partial}{\partial x}\right), \feq the  Lagrangian
(\ref{sg}) takes  the form 
\bea\lb{sg1}
{\cal{L}}=\frac{\omega_{t}}{2}\hat \partial_{\nu}\theta\hat
\partial^{\nu}\theta +\frac{r^{2} \omega_{s}}{2}\left(
\partial_{\nu}\theta \partial^{\nu}\theta+ 2 \cos
(\Phi-\theta)\partial_{\nu}\theta
\partial^{\nu}\Phi +\partial_{\nu}\Phi
\partial^{\nu}\Phi\right)\nonumber\\
 + 4r^{2}  K_{p}\left( \cos\theta +\cos\Phi  -
\frac{1}{2}\cos(\Phi-\theta) -\frac{3}{2}\right)-V_{c}(\Phi-\theta). \eea
Generalizing the Lagrangian (\ref{sg1}) one obtains a class of
models admitting solitonic solutions with fixed speed.

The models have the form of a sine-Gordon-like coupled system of
two scalar fields and will be functionally parametrized by two
coupling functions $F,G$. In the Minkowski spacetime notation the
Lagrangian is given by \bea\lb{sg2} {\cal{L}}_{g}
&=&\frac{a}{2}\hat
\partial_{\nu}\theta\hat
\partial^{\nu}\theta +\frac{b}{2} \partial_{\nu}\theta
\partial^{\nu}\theta+ F(\Phi-\theta)\partial_{\nu}\theta
\partial^{\nu}\Phi +\nonumber\\
&+&\frac{b}{2} \partial_{\nu}\Phi \partial^{\nu}\Phi + K\left(
\cos\theta +\cos\Phi -2 \right)+ G(\Phi-\theta) , \eea where
$a, b,K$ are positive  constant  parameters and the coupling
functions $F(\Phi-\theta)$ and $G(\Phi-\theta)$ are arbitrary {\sl
even} functions of $\Phi-\theta$ and $G$ satisfies $G(0)=0$.
Considering only travelling wave
solutions $\Phi(z),\theta(z)$ (with $z=x+vt$) we obtain the 1D
reduced Lagrangian which is the generalization of the one given in
(\ref{lagra}): \beq\lb{sg2a} L= \frac{1}{2} \widetilde J
(\theta')^{2}+ \frac{\widetilde \mu  }{2}\left[ (\theta')^{2}+2 F
\theta'\Phi'+(\Phi')^{2} \right] +K\left[  \cos\Phi + \cos
(\theta)- 2\right]+ G, \feq where $$ \widetilde J=
a(v^{2}/c^{2}_{t}-1)\ , \ \ \widetilde \mu= b(v^{2}/c^{2}_{s}-1) \
. $$ Taking into account that because of the parity of the
coupling functions $F$ and $G$ we have $F'(0)=G'(0)=0$, it is not
difficult to show that for $\Phi=\theta$ the  equation of motion
stemming from the action(\ref{sg2a}) admits the SG soliton
(\ref{fe3}) travelling at speed
$v=c_{t}<c_{s}$.

A further generalization, which we will not investigate in detail
in this paper,  can be obtained considering generically a
self-interacting scalar field $\Theta$ in 2D Minkowski spacetime,
\beq\lb{si} {\cal L}=
\frac{1}{2}\partial_{\nu}\Theta\partial^{\nu}\Theta - V(\Theta),
\feq whose potential $V$ is such that field equations for $\Theta$
admit relativistic soliton solutions (apart from the SG system,
the model contains as a particular case e.g the $\phi^{4}$
interaction model). A general Lagrangian describing two
interacting scalar fields $\Theta,\Sigma$ for which the speed
selection mechanism for the soliton applies is given by
\beq\lb{sg3a} \begin{array}{ll} {\cal{L}}_{g} = \frac{a}{2}\hat
\partial_{\nu}\Theta\hat \partial^{\nu}\Theta +\frac{b}{2}
\partial_{\nu}\Theta \partial^{\nu}\Theta+
F(\Theta-\Sigma)\partial_{\nu}\Theta
\partial^{\nu}\Sigma +\frac{b}{2}\partial_{\nu}\Sigma
\partial^{\nu}\Sigma \\
 \ \ \ \ - V(\Theta) - V(\Sigma) - G(\Theta-\Sigma) \ ,
 \end{array}
\feq where as above $F(\Theta-\Sigma)$ and $G(\Theta-\Sigma)$ are
arbitrary even functions of $(\Theta-\Sigma)$.

\section{Minimal model}
\lb{S5}
The Lagrangian (\ref{sg2}) corresponds to a system of two coupled
sine-Gordon models. It is  of interest to see if the properties
concerning the selection of the soliton speed hold also when the
interaction terms are switched off, i.e when the coupling
functions $F,G$ are identically zero. Moreover in the completely
non-interacting case,  the system will be integrable and we will
have complete control on the general solutions.

For $F=G=0$ the Lagrangian (\ref{sg2}) becomes \beq\lb{sg3}
{\cal{L}}_{m} =\frac{a}{2}\hat \partial_{\nu}\theta\hat
\partial^{\nu}\theta +\frac{b}{2} \partial_{\nu}\theta
\partial^{\nu}\theta+ \frac{b}{2} \partial_{\nu}\Phi
\partial^{\nu}\Phi + K\left( \cos\theta +\cos\Phi -2 \right). \feq
The field equations describe a decoupled system for $\theta$ and
$\Phi$,
\beq\lb{s3}
\frac{a}{b}\hat \partial_{\nu}\hat \partial^{\nu}\theta+
\partial_{\nu} \partial^{\nu}\theta=
-\frac{K}{b}\sin\theta,\quad \partial_{\nu}
\partial^{\nu}\Phi=-\frac{K}{b}\sin\Phi.
\feq
The theory contains two upper bound speeds, $c_{s},c_{m}¥$ for the 
propagation of  respectively $\Phi$- and $\theta$-waves. $c_{m}$ is 
given by
\beq\lb{speeds} c_{m}=c_{t}c_{s} \sqrt{\frac{a+b}{a c^{2}_{s}+b c^{2}_{t}}}.
\feq
Notice that $c_{s}>c_{t}$ implies necessarily also $c_{s}>c_{m}$.

Because the two scalars are decoupled we can look for
traveling wave solutions of different speeds $v,\bar v$ for the two
fields: $\Phi(z)=\Phi(x+vt),\,$ $\theta(\bar z)=\theta(x+\bar vt)$.
In this way we  we end up with
a decoupled system of differential equations,
\beq\lb{f7} \left(\frac{ v^{2}}{c^{2}_{s}}-1\right)\Phi''=
-\frac{K}{b} \sin \Phi,\quad \left[\frac{a}{b}\left(\frac{{\bar
v}^{2}}{c^{2}_{t}}-1\right)+ \left(\frac{{\bar
v}^{2}}{c^{2}_{s}}-1\right)\right]\ddot\theta= -\frac{K}{b} \sin
\theta, \feq
where the prime and the dot denote derivation with
respect to $z$ and $\bar z$ respectively. The general solution of
the previous system is given by two sine-Gordon solitons of type
(\ref{fe3}) with speeds  $v\le c_{s}$ and $\bar v<c_{m}<c_{s}$, 
respectively for the fields  $\Phi$ and $\theta$. Only
imposing $\theta=\Phi$ we have velocity selection. In fact
$\theta=\Phi$ implies  $v=\bar v=c_{t}$ Notice that the converse
is not true. $v=c_{t}$ does not  necessarily imply $\theta=\Phi$
or $v=\bar v$. For $v=c_{t}$ and $\theta\neq \Phi$
 we will have solitons  in $\theta$ and $\Phi$ that differ in speed
or, if the speed is the same, in the coordinate of the center of the
soliton ( we will have two identical solitons with non-zero relative
phase).

\section {Lorentz Symmetry}
\lb{S6}
It is well known that usual  SG field theory, described by the 2D
action
\beq\lb{usg} S=\int  \left[\frac{(\phi_{t})^{2}¥}{2 c_{0}^{2}}-
\frac{1}{2}(\phi_{x})^{2}¥+ \omega_{0}(\cos\phi-1)\right] \, dx dt \ ,
\feq is invariant under 2D Lorentz transformations (boosts)
\beq\lb{boost} x=\gamma(x'+vt'),\quad t=
\gamma(t'+\frac{v}{c_{0}^{2} }x'),\quad
\gamma=\frac{1}{\sqrt{1-\frac{v^{2}}{c_{0}^{2} }}}, \feq where
$c_{0}$ plays the role  of the speed of light in vacuum. The
Lorentz symmetry of the action (\ref{usg}) is essential for
establishing the soliton/relativistic particle analogy and allows
one to use for SG solitons the concepts of relativistic
kinematics. For instance, Lorentz symmetry allows us to generate,
using a boost (\ref{boost}), a SG soliton propagating with speed
$v$ \beq\lb{sgs} \phi(x+vt)= 4\arctan
\left[e^{\omega_{0}\gamma(x+vt)}\right], \feq from the static
solution $\phi_{0}(x)= 4\arctan [e^{\omega_{0}x}]$. Another
consequence of the Lorentz symmetry of the theory is the existence
of a maximum propagation speed, $v_{max}=c_{0}$ for the solitons.
To make the Lorentz symmetry  more evident we might write the action
(\ref{usg}) using the Minkowskian notation of Eqs. (\ref{fe11}).

On the other hand, the two-fields generalized SG theory
(\ref{sg2}) we are discussing in this paper is {\it not } Lorentz
invariant. This is due the fact that in the Lagrangian (\ref{sg2})
appear not a single upper bound speed $c_{0}$ but 
two of them $c_{s}, c_{m}$. It follows that we
have the choice of defining the boosts (\ref{boost}) either in
terms of $c_{s}$ or $c_{m}$. In the first case the Lorentz
symmetry is broken by the kinetic term proportional to $a$, in the
second case by the kinetic terms proportional to $b$ in the action
(\ref{sg2}). Because $c_{m}<c_{s}$ it is natural to  consider
boosts of the first kind, so that it is the $a$-term that breaks
the Lorentz symmetry.

From the physical point of view, the breaking of the Lorentz
symmetry can be easily understood if one realizes that our system
can be thought as a  medium with two different  speeds 
for the propagation of sound waves.

It is obvious that the speed selection effect described in
previous sections is strongly related to breaking of the Lorentz
symmetry. In fact if in Eq. (\ref{sg2}) we set  $a=0$ we have no
selection for the speed of the soliton and Lorentz symmetry is not
broken. Let us therefore investigate in detail the relationship
between breaking of Lorentz symmetry and soliton speed selection
mechanism, starting from the minimal model (\ref{sg3}).

\subsection{The minimal case}

In the minimal case, the fields, $\Phi$ and $\theta$ do not
interact one with the other. This means that the field theory has
two completely decoupled sectors. The field equations (\ref{s3})
can be written as follows, \beq\lb{t1}
\frac{1}{c_{m}^{2}}\theta_{tt}-\theta_{xx}=-\frac{K}{a+b}
\sin\theta,\quad \frac{1}{c_{s}^{2}}\Phi_{tt}-\Phi_{xx}=
-\frac{K}{b} \sin\Phi, \feq where $c_{m}$ is as in Eq.
(\ref{speeds}). Notice that  we have three fundamental speeds in
the theory, viz. $c_{m},c_{t},c_{s}$ with $c_{m}<c_{t}<c_{s}$.
Here $c_{m}$ and $c_{s}$ are the maximal speeds of propagation
respectively for $\theta$-  and $\Phi$-waves. The physical meaning
of $c_{t}$ will be clear later.

Because of the decoupling, the two sectors of the field theory are
invariant under two different groups of Lorentz transformations.
Using a notation in which the dependence of a boost from the upper
bound speed is evident, we will denote with $\Gamma(c_{0})$ the
boost in (\ref{boost}). Whereas the field equation (\ref{t1}) for
$\theta$ is invariant under boosts $\Gamma(c_{m})$, that for
$\Phi$ is invariant under $\Gamma(c_{s})$. The previous symmetry
properties have a simple physical consequence: the existence of
solitonic excitation for the fields $\theta$ and $\Phi$  which
propagate at two independent speeds $\bar v,v$ (see section \ref{S5}).
The $\Phi$-soliton, $\Phi(x+ v t)$ can be obtained boosting the
static soliton $\Phi(x)$  with\ $\Gamma(c_{s})$, whereas the
$\theta$-soliton, $\theta(x+\bar v t)$ can be obtained boosting
the static soliton $\theta(x)$  with\ $\Gamma(c_{m})$. The
solutions  can be written in an Lorentz invariant form introducing
the covariant vectors \beq\lb{cov} k^{\mu}=(k,\omega), \quad  \bar
k^{\mu}=(\bar k,\bar \omega),\quad x^{\mu}=(c_{s}t, x),\quad \bar
x^{\mu}=(c_{m}t, x). \feq Propagating solutions are function of
the Lorentz scalars, $$ \Phi = \Phi(k_{\mu}x^{\mu},
k_{\mu}k^{\mu}) \ \ , \ \ \  \theta=\theta(\bar k_{\mu}\bar
x^{\mu}, \bar k_{\mu}\bar k^{\mu}) \ . $$

The condition $\theta=\Phi$ couples the two otherwise decoupled
sectors and breaks Lorentz invariance. In fact $\theta=\Phi$
requires $v =\bar v$, whereas Lorentz invariance would require
$\Gamma(c_{s})=\Gamma(c_{m})$, which is  manifestly impossible for
$c_{s}\neq c_{m}$. As shown in Section \ref{S5}, the condition
$\theta=\Phi$ fixes the  speed of the soliton so that we have $v
=\bar v=c_{t} $. The soliton propagating at this speed cannot be
obtained by boosting a static solution. This is  also evident
considering that  Eqs. (\ref{f7}) do not allow for  static
solutions with $\Phi=\theta$.

From a physical point of view the mechanism described above can be
seen as a speed selection effect generated by breaking of the Lorentz
symmetry. We have  a media with two different maximal propagation
speeds, $c_{s}$, $c_{m}$ that can be identified as the propagation
speed of $\Phi$- and $\theta$-sound waves. Furthermore, we have a constraint,
$\Phi=\theta$ which breaks the Lorentz symmetry and singles out a
solitonic perturbation which can propagate only at the fixed speed
$c_{t}$.

\subsection{Interacting models}

Until now we have considered only the minimal, non interacting
model (\ref{sg3}). Passing to consider the interacting model
(\ref{sg2}) there is only one main difference: as now the fields
$\Phi,\theta$ are coupled to each other, only one single boost
$\Gamma(c_{s})$ can be defined, and the Lorentz symmetry is broken
already at level of the action (\ref{sg2}). Being mutually
coupled, travelling wave solutions for $\theta,\Phi$ now have to
propagate with the same velocity $v$. Using the  covariant
notation of Eq. (\ref{cov}) they take the form \beq\lb{so4}
\Phi=\Phi(k_{\mu}x^{\mu}, k_{\mu}k^{\mu}, \omega^{2}c_s^2/c_m^2 -
 k^{2}) \ , \ \ \theta = \theta(k_{\mu}x^{\mu}, k_{\mu}k^{\mu},
\omega^{2}c_s^2/c_m^2 - k^{2}). \feq Owing to the dependence on
the non-Lorentz invariant term $(\omega^{2}c_s^2/c_m^2 - k^2)$ the
solutions are not Lorentz-invariant. Although the boosts
$\Gamma(c_{s})$ are not a symmetry of the Lagrangian (\ref{sg2})
they can be used as
solution-generating transformations. A solitonic solution
travelling with speed $v$ can be obtained boosting a static
solution. 
Similarly to what happens in the minimal case, the condition
$\theta=\Phi$ breaks this solution-generating symmetry. The
$\theta=\Phi$ solution cannot be generated by boosting a static
solution satisfying the constraint, so that the velocity $v=c_{t}$
for the soliton is selected.

\section{ Discrete symmetries and topological classification}
\lb{S7}
The internal (discrete)
symmetries of our model are very important both for understanding
the peculiarities of the fixed-speed $\theta=\Phi$ solution and
for the topological classification of the solitons.

We are dealing with a 2D field theory with broken Lorentz
symmetry, for which the internal symmetry group of the Lagrangian
(\ref{sg2}) is not necessarily the same as that of the reduced
action (\ref{sg2a}) describing travelling wave solutions. We are
only interested in the symmetries relevant for the solitonic
solutions, namely in the symmetries of the reduced Lagrangian
(\ref{sg2a}).

For $\widetilde J\neq 0$ and for generic coupling functions $F,G$
the internal symmetry group of the model -- denoted as $\cal{G}$
in the following -- and the residual symmetry of the vacua --
denoted as $\cal{H}$ below -- are the same as those of the simple
SG system (\ref{usg}). We have ${\cal G}= {\bf Z}\times {\bf
Z}_{2}$, where ${\bf Z}$ is realized as discrete translations
$\Phi\to \Phi+ 2\pi n,\,\theta\to \theta+ 2\pi n$, whereas ${\bf
Z}_{2}$ is the inversion $\theta \to -\theta,\, \Phi\to -\Phi$.
The vacua $\Phi=\theta =2m\pi$ are invariant under an inversion
followed by a translation of $4\pi m$. The residual symmetry group
of the vacua is therefore ${\cal H}= {\bf Z}_{2}$.

For $\widetilde  J=0$ the group ${\cal G}$  acquires an additional
${\bf Z}_{2}$ factor: ${\cal G}= {\bf Z}\times {\bf Z}_{2}\times
{\bf Z}_{2}$. This third factor comes from the invariance of the
Lagrangian (\ref{sg2a}) under the transformation \beq\lb{f5}
\theta \longleftrightarrow \Phi \ . \feq The group ${\cal H}$
acquires an additional factor as well: ${\cal H}= {\bf
Z}_{2}\times {\bf Z}_{2}$. Notice that the speed-selecting
constraint leaves invariant both $\cal{G}$ and
$\cal{H}$. In fact  
the submanifold $\Phi = \theta$ is  selected by the constraint.

The solitonic solutions of the model can be therefore classified
in two classes, which are mapped one into the other by the
transformation (\ref{f5}). The solitonic solutions with
$\theta=\Phi$ are self-dual, i.e a fixed point of the
transformation (\ref{f5}).

The previous discussion holds true for generic coupling functions
$F,G$. For  particular choices of $F$ and $G$ we can have  bigger
 $\cal{G}$ and $\cal{H}$ groups. This happens for instance when $F$
and $G$ are periodic functions. In this case  $\cal{G}$ is the
product of two ${\bf Z}$ groups times some ${\bf Z}_{2}$ factors.
However, the $\theta=\Phi$ condition breaks invariance under one
of the two ${\bf Z}$ groups and we are left with the same symmetry
groups $\cal{G}$ and $\cal{H}$ as above.

The discrete symmetries of SG-like field theories are  also of
fundamental importance for the topological classification of the
solitonic excitations of the theory. The admissible number of
solitons of the SG model  is determined by the number of ways in
which the points $x=\pm \infty$  can be
mapped into the different vacua of the model.

We have therefore a one-to-one correspondence between solitons and
the elements of the homotopy group $\pi_{0}({\cal G}/{\cal H})$.
It is also well-known that field theories that admit topological
solitonic solutions have conserved currents corresponding to
conserved topological charges, which can be identified with the
winding numbers of the soliton.

Using the previous results one can easily calculate the homotopy
group $\pi_{0}$ in the case of generic coupling functions $F,G$.
For $\widetilde J\neq 0$ this is given by $$ \pi_{0} ({\cal
G}/{\cal H}) \ =\ \pi_{0} ({\bf Z} \times {\bf Z}_{2})/{\bf Z}_{2}
\ = \ {\bf Z} \ . $$ The same result holds true when $\widetilde
J\neq 0$:
$$ \pi_{0} ({\cal G}/{\cal H}) \ = \ \pi_{0} ({\bf Z} \times {\bf
Z}_{2} \times {\bf Z}_{2})/({\bf Z}_{2}\times {\bf Z}_{2}) \ = \
{\bf Z} \ . $$ As expected the solitonic solution of the model can
be labeled by an integer $n$, its topological charge.

For periodic coupling functions the homotopy group $\pi_{0}$ of the
model becomes ${\bf Z}\times {\bf Z}$ and we have two topological
charges $(n,m)$ associated to the soliton. However, the condition
$\theta=\Phi$ reduces the symmetry group and we are left again with
$\pi_{0}={\bf Z}$ and a single topological charge $n$.

\section{Stability}
\lb{S8}
The stability of the solutions we have considered in the previous
section is a crucial requirement for our speed selection mechanism.
First, being the dynamics non linear we expect high sensitivity of
the solutions to the initial conditions. If the $\theta=\Phi$
solution is not stable,  small
perturbations in the initial conditions may disrupt the mechanism.
Second, even though a soliton with the given speed is generated, it
may be unstable under perturbation and decay in a finite time.

In the previous section we have seen that the fixed-speed
solitonic SG soliton (\ref{fe3}), although resulting from breaking
of the Lorentz symmetry, share many feature with the usual SG
solitons: they are localized, finite energy  solutions of the same
ordinary differential equation, they allow for the same
topological classification and associated topological charges.
These features strongly  suggest that the fixed-speed SG solitons we are
discussing in this paper are stable solutions, in the same way as
relativistic SG solitons are.

The analysis of the stability of the solutions of the model
(\ref{sg2}) is rather complicate.
In this paper we will restrict our investigations to  the classical stability of
the speed-selecting solution $\theta=\Phi$ under  linear perturbations.
To begin with, we will  briefly summarize well-known results
about the classical stability of the Solitons  of the simple SG model
(\ref{usg}).

To discuss the stability one first uses Lorentz invariance to
chose the reference frame in which the soliton (\ref{sgs}) is
static and then considers small perturbations around the static
solution, 
\beq\lb{s1} \phi(x,t)=\phi_{0}(x)+\psi(x,t), \quad
\psi<<1. \feq 
Inserting this position in the field equations,
considering only the linear approximation and separating the
variables in the perturbation,
i.e setting $\psi(x,t)=\Psi(x) e^{i\omega t}$, one obtains an
equation for $\Psi$, which has the form of a time-inde\-pendent
Schrodinger equation 
\beq\lb{se}
T\Psi=\left(\frac{\omega}{c_{0}}\right)^{2}\Psi,\quad T=
-\frac{d^{2}}{dx^{2}}+\omega_{0}^{2}(1-2{\rm sech}^{2}x). \feq
Classical stability requires  the eigenvalues of the operator $T$ to
be real and non-negative. This is guaranteed if the operator $T$ is
self-adjoint and non-negative. These two properties of the operator
$T$ given in Eq. (\ref{se}) can be shown in different ways.

The most elegant way is  to use supersymmetric (SUSY)
factorization \cite{witten1,cooper,cooper1}. One can find a
superpotential $W(x)$ such that $T=A^{+}A$ with $A=(d/dx)-W(x)$,
$A^{+}=-(d/dx)-W(x)$, from which immediately follows that $T$ is
self-adjoint, whereas the SUSY algebra implies that the spectrum
of $T$ is non-negative. Being $T$ a  SUSY Hamiltonian, the
Schrodinger equation (\ref{se}) can be explicitly solved. It turns
out that the spectrum has a single discrete mode $\psi_{0}$ for
$\omega=0$, and a continuum part $\psi_{k}$ for
$\omega>\omega_{0}c_{0}$ \cite{daupey}.

The discrete mode corresponds to soliton translations, whereas the
continuos modes are not reflected by the potential in Eq.
(\ref{se}) and describe soliton deformations. The set
$\{\psi_{0},\psi_{k}\}$ is a complete, orthonormal basis and can
be therefore used to expand a generic perturbation $\psi(x,t)$.

Let us now consider our generalized SG model (\ref{sg2}). To
simplify the calculation we will first consider a vanishing
coupling function $F$, later we will discuss the most general
case. For $F=0$, introducing the new dimensionless variables
$t'=\omega_{0} c_{s} t$ and $x'= \omega_{0} x$, with
$\omega_{0}= K/b$ and the dimensionless parameter $\alpha=a/b$,
the field equations stemming from the Lagrangian (\ref{sg2}) are
\bea\lb{s4} &&\Phi_{t't'}-\Phi_{x'x'}=-\sin\Phi
+\frac{1}{\omega_{0}^{2}}\frac{\partial G}{\partial \Phi}\nonumber\\
&&\left(1+ \alpha\frac{c^{2}_{s}}{c^{2}_{t}}\right)\theta_{t't'}-
\left(1+ \alpha\right)\theta_{x'x'}=-\sin\theta
+\frac{1}{\omega_{0}^{2}}\frac{\partial G}{\partial \theta}.
\eea
We want to investigate the classical stability of the $\theta=\Phi$
soliton solution of the previous equations,  which propagates at
fixed speed $v=c_{t}$. It is therefore appropriate to choose a frame
in which this soliton is at rest. This can be easily done by
performing in Eqs. (\ref{s4}) a boost of speed $c_{t}$,
\beq\lb{boost1}
y=\gamma(x'+\frac{c_{t}}{c_{s}} t'),\quad
\tau=\gamma(t'+\frac{c_{t}}{c_{s}} x'),\quad \gamma= \frac{1}{\sqrt{1-
(\frac{c_{t}}{c_{s}})^{2}}}.
\feq
In the $(\tau,y )$ frame, Eqs. (\ref {s4}) read
\bea\lb{s5}
&&\Phi_{\tau\tau}-\Phi_{yy}=-\sin\Phi
+\frac{1}{\omega_{0}^{2}}\frac{\partial G}{\partial \Phi}\nonumber\\
&&\theta_{\tau\tau}-\theta_{yy}+
 \alpha\left(1+
 \frac{c^{2}_{s}}{c^{2}_{t}}\right)\theta_{\tau\tau}+
 2\alpha\frac{c_{s}}{c_{t}}\theta_{\tau} \theta_{y} =-\sin\theta
+\frac{1}{\omega_{0}^{2}}\frac{\partial G}{\partial \theta}.
\eea
Notice that the third and fourth term of the l.h.s. of the second
equation are responsible for the breaking of Lorentz symmetry.
Seen in the $(\tau,y)$ frame, the solitonic solution of fixed
speed $c_{t}$ is static; it is given by  \beq\lb{ss}
\Phi_{0}(y)=\theta_{0}(y)=\arctan e^{y}. \feq The classical
stability of the previous solution is investigated by considering
small perturbations $\sigma,\chi$ around the static solution,
\beq\lb{pert} \Phi(y,\tau)=\theta_{0}(y)+\sigma(y,\tau),\quad
\Theta(y,\tau)=\theta_{0}(y)+\chi(y,\tau). \feq At first order in
the perturbation Eqs. (\ref{s5}) give, \bea\lb{s6}
&&\sigma_{\tau\tau}-\sigma_{yy}=-(\cos\theta_{0})\sigma
+B(\sigma-\chi)\nonumber\\
&&\chi_{\tau\tau}-\chi_{yy}+
 \alpha\left(1+
 \frac{c^{2}_{s}}{c^{2}_{t}}\right)\chi_{\tau\tau}+
 2\alpha\frac{c_{s}}{c_{t}}\theta_{\tau} (\theta_{0})_{y} =-\sin\theta
-B(\sigma-\chi), \eea where $B$ is a constant given by
$B=(1/\omega_{0}^{2})(\partial^{2}G/\partial\Phi^{2})(0).$

For generic values of the parameter $\alpha$, the system
(\ref{s6}) is very hard to solve. For
$\alpha<<1$ we can treat the terms proportional to $\alpha$ as a
perturbation. At zeroth order in the perturbative expansion we can
neglect them. In this case we can write Eqs. (\ref{s6}) in
operator form, introducing the Schrodinger-like operator $R$:
\beq\lb{L} R\left(\begin{array}{c}
\sigma \\
\chi\\
\end{array}\right)=-\frac{\partial^{2}}{\partial \tau}
\left(\begin{array}{c}
\sigma \\
\chi\\
\end{array}\right).
\feq The operator $R$ can be diagonalized  by the linear
transformation $\phi_{1} =\sigma-\chi$, $\phi_{2} =\sigma+\chi$
\beq\lb{L1} R=\left(\begin{array}{cc}
-\frac{\partial^{2}}{\partial y^{2}}+V_{1}(y)&0 \\
0&-\frac{\partial^{2}}{\partial y^{2}}+V_{2}(y)\\
\end{array}\right).
\feq where the potentials $V_{1}$ and $V_{2}$ are \beq\lb{pot}
V_{1}=(1-2 {\rm sech}^{2}y),\quad V_{2}=(1+2B-2{\rm sech}^{2}y).
\feq Separating the variables in the perturbations, i.e setting
$\phi_{1}(y,\tau)=\phi_{1}(y)e^{i\omega_{1}  \tau}$, $
\phi_{2}(y,\tau)=\phi_{2}(y)e^{i\omega_{2}  \tau}$, equation
(\ref{L}) becomes a time-independent Schrodinger-like equation.

There are two particular values of  the parameter $B$ which
correspond to SUSY models, $B=0$ and $B=3/2$. In both cases the
operator $R$ is self-adjoint and non-negative, implying stability
of the background solution. The $B=0$ case corresponds to the
minimal, decoupled case discussed in Sect. \ref{S5}.  The diagonal
elements $R_{1}$ and $R_{2}$ of the operator $R$ of Eq. (\ref{L1})
 corresponding respectively to the perturbations $\phi_{1},\phi_{2}$,
 are equal and coincide with the previously discussed operator $T$ of
 Eq. (\ref{se}). The superpotential is also the same and is given by
 $W_{1} =W_{2}=\tanh y$.
$R_{1}$ and $R_{2}$ have therefore the same spectrum, which
coincides with that of the operator $T$.

For $B=3/2$, the operators $R_{1}$ and $R_{2}$, although not
equal, correspond both to SUSY Hamiltonians. $R_{1}$ is equal to
the operator $T$ of Eq. (\ref{se}) whereas $R_{2}$ correspond to a
different SUSY Hamiltonian (see e.g. Ref. \cite{cooper, cooper1}).
The superpotential is given by $W_{2}=2 \tanh(y)$. The spectrum of
$R_{2}$ has two discrete modes with eigenvalues $\omega_{2} =0$
and $\omega_{2} =(3/4) \omega_{0 }c_{s}$. The continuum part of
the spectrum is located at $\omega_{2}\ge 4\omega_{0}c_{s}.$

For generic $B$  the operator $R_{1}$, being independent from $B$,
corresponds always to the SUSY Hamiltonian discussed above. On the
other hand the operator $R_{2}$ does not seem  to correspond to a
SUSY Hamiltonian. From the general form of the potential $V_{2}$,
see Eq. (\ref{pot}), one can easily infer that we have a discrete
spectrum for $\omega_{2}< (1+2B)\omega_{0}c_{s}$, whereas we have
a continuum spectrum for $\omega_{2}\ge (1+2B)\omega_{0}c_{s}$.
However, we do not have any argument to show that the eigenvalues
of the discrete spectrum of $R_{2}$ are non negative. In the
following we will consider only the cases $B=0$, but our
considerations can be easily extended to the case $B=3/2$.

Let us now consider the first order in the perturbation expansion
in the parameter $\alpha$, i.e let us consider the system in the
complete form (\ref{s6}). For $B=0$ the first equation of the
system can be written as \beq\lb{s11}
\sigma_{\tau\tau}-\sigma_{yy}=-(1-2 {\rm sech}^{2}(y) \sigma. \feq
Comparing with Eq. (\ref{se}), we immediately realize that this
corresponds to the operator $T=R_{1}$. It follows immediately that
our fixed speed solitonic solution (\ref{ss}) is stable for small
perturbation of the field $\Phi$. After some manipulations the
equation for the perturbation $\chi$ in (\ref{s6}) (and $B=0$) can
be written as \beq\lb{s12} C\chi_{\tau\tau}-\chi_{yy}+
 D ({\rm sech} y) \chi_{\tau}+ (1-2{\rm sech}^{2}y)\chi=0,
\feq where $C=1+\alpha[1+(c_{s}/c_{t})^{2}],\,
D=\alpha(c_{s}/c_{t})$. The previous equation differs from the
equation (\ref{s11}) in two points. First, the term
$\chi_{\tau\tau}$ is multiplied by $C$. This is rather harmless,
it just means that the eigenvalues of the operator $T$ have to be
rescaled by the positive factor $C$. This term cannot change the
semi-positivity of the eigenvalues. Second, there is an
$y$-dependent dissipative term. This term becomes negligible for
$y\to\pm \infty$. In fact in this limit Eq. (\ref{s12}) gives
$$ C\chi_{\tau\tau}-\chi_{yy}+\chi=0,$$
which has normal mode solutions $\chi=Ae^{i(ky+\omega\tau)}$ with
dispersion relation $C\omega^{2}= k^{2}+1$.

The damping term is maximized near the minimum of the potential
$V(y)$, $y=0$. Near $y=0$ the perturbation $\chi$ has a damping
term $\chi\propto e^{-\beta \tau}$, where $\beta$ is  some
constant. It is interesting to notice in this context that the
Lorentz symmetry breaking terms act on the perturbation  as
damping term and as rescaling of the eigenvalues of the operator
$T$.

The equation (\ref{s12}) can be solved by expanding $\chi$ in
terms of the eigenfunctions $\{\psi_{0},\psi_{k}\}$ of the
operator $T$ which represent a complete orthonormal set,
\beq\lb{s14} \chi(y,\tau) \ = \ \psi_{0}(y) h(\tau) \ + \
\int^{\infty}_{-\infty} \psi_{k}(y) g_{k}¥(\tau) \, dk \ . \feq
Inserting this equation into (\ref{s12}), using the orthonormal
character relations, and the fact that $\{\psi_{0},\psi_{k}\}$ are
eigenfunction of the operator $T$ corresponding to eigenvalues $0$
and $\omega^{2}$, one obtains after some manipulations the
equations for $h$ and $g_{k}¥$; these turn out to be \bea\lb{s15}
&&C\frac {d^{2}h}{d\tau^{2}}+ 2\pi D \frac{dh}{d\tau}=0,\nonumber \\
&&C\frac {d^{2}g_{l} }{d\tau^{2}}+ D\int_{-\infty}^{\infty}dk\,dy
[{\rm sech}(y) {\psi^{*}_{l}\psi_{k}} \frac{dg_{k}¥}{d\tau}]+
\omega^{2}g_{l} =0. \eea

The equation for $h$ is immediately solved to give \beq\lb{s16}
h(\tau)= Pe^{-\beta \tau}+Q, \feq where $P,Q$ are integration
constants and $\beta= 2\pi D/C$. From Eq. (\ref{s16}) it follows
that after a transition time, only the pure translation mode
$\psi_{0}$ of the SG soliton will survive (when $Q\neq 0$); or
that this mode will be completely damped to zero and only the
static background SG solution will survive (for $Q=0$). In both
cases, this shows  stability of our fixed speed soliton under
perturbations of the discrete spectrum.

The equation for the functions $g_{l} (\tau)$ in Eq. (\ref{s15})
is more involved.  It is a integro-differential equation that
cannot be solved in closed form. However, from the general
structure of the differential equation, it is evident that the
solutions will always have an oscillatory part and a damping term
$\sim e^{-\beta_{l}\tau}$.

Qualitatively the behavior of $g$ will be therefore similar to the
previously described one for $h$. After a transition time, the
normal modes corresponding to the continuum spectrum of the
operator $T$ will be either completely damped or will survive as
simple soliton deformations. This provides strong
arguments for the stability of our solitonic
solution for perturbations of the continuum spectrum.

Until now we have considered  explicitly  stability when the
coupling function $F$ vanishes. Our discussion can be easily
extended to the case when $F\neq 0$. Looking at the Lagrangian
(\ref{sg2}) one easily realizes that the terms introduced in this
way are invariant under Lorentz transformations $\Gamma(c_s)$. In
the perturbation approach these terms have two effects. They
change the value of the parameter $C$ in Eq. (\ref{s12}), and
introduce a term proportional to ${\rm sech}^{2}y(\sigma-\chi)$ in
both Eqs. (\ref{s6}). The modification of $C$ does not change the
qualitative outcome of our previous discussion, whereas the
modified Eqs. (\ref{s6}) can be analyzed using the same method we
have previously used.

\section{Conclusions}
\lb{S9}

In this paper we have proposed a mechanism for fixing the velocity
of relativistic solitons. The proposal has been first elaborated
for a molecular chain model of double coupled
 pendulums. It has been then generalized to a full class of 2D
field theories of the sine-Gordon type.

From a phenomenological point of view, the model allows one to
select the speed of a SG soliton just by tuning the elastic
coupling constants and the kinematical parameters that support the
transverse sound wave. Thus, the selection speed mechanism could
be at work whenever SG solitons appear with a given speed -- and
could be useful to formulate models whenever one needs to explain
phenomena of this type, or however where the soliton speed plays a
special role and should be tuned.
Moreover, our selection mechanism could be also used to devise and 
realize non-linear media where SG solitons travel at a given fixed 
speed.

Our results are also relevant from a fundamental,
field-theoretical point of view. In fact we have shown that the
speed selection mechanism is deeply  related both with the
existence of some conditionally conserved quantity
and to the breaking of the Lorentz
symmetry of the usual SG models. The most striking and highly non
trivial result we have obtained in this context is that the
characterizing features of relativistic SG solitons -- that is,
finiteness and localization of energy, topological classification
and stability -- may be still preserved even if the Lorentz
symmetry is broken and a soliton of a given speed is selected.

We have shown that the solitonic excitation with fixed speed of
our model are classically stable and have topological conserved
charges. In particular it turns out that the Lorentz symmetry
breaking terms act on the perturbations around the fixed speed
soliton as a damping term.

This result is particularly intriguing in view of the analogy
between relativistic particles and SG solitons. It can be seen as
a strong indication that relativistic nature of the particle is
not a necessary condition for the analogy to hold. In particular
one can speculate about the existence of SG solitons which behave
as non-relativistic (e.g Newtonian) particles.


\begin{thebibliography}{99}

\bibitem{russell}
J. Scott Russell,  ``Report on waves'', {\it Proc. of the British
Association for the Advancement of Science},  London, (1845).

\bibitem{kdv}
D. J. Korteweg and G. de Vries, Phil. Mag. {\bf 39}, 422 (1895).

\bibitem{zk}
N. J. Zabusky and M. D. Kruskal, Phys. Rev. Lett. {\bf 15}, 240 (1965).

\bibitem{daupey}
Th. Dauxois and M. Peyrard, {\it Physique de solitons}, Editions
CNRS (Paris) 2004; {\it Physics of Solitons}, Cambridge UP
(Cambridge) 2006.
\bibitem{dodd}
R.K. Dodd, J. C. Eilbeck, J. D. Gibbon, H.C. Morris,
{\sl Solitons and Nonlinear Wave Equations}, Academic Press, London,
New York (1982).
\bibitem{rubinstein}
J. Rubinstein, J. Math. Phys. {\bf 11}, 258 (1970).

\bibitem{Dav} A.S. Davydov, {\it Solitons in Molecular Systems},
Kluwer (Dordrecht) 1981.

\bibitem{Eng} S.W. Englander, N.R. Kallenbach, A.J. Heeger, J.A.
Krumhansl and A. Litwin, {\it Proc. Nat. Acad. Sci. USA} {\bf 77}
(1980), 7222-7226.


\bibitem{YakuBook} L.V. Yakushevich, {\it Nonlinear Physics of
DNA}, Wiley (Chichester) 1998; second edition 2004.

\bibitem{GRPD} G. Gaeta, C. Reiss, M. Peyrard and Th. Dauxois,
{\it Rivista del Nuovo Cimento} {\bf 17} (1994) n.4, 1--48.

\bibitem{PeyNLN} M. Peyrard, {\it Nonlinearity} {\bf 17}  R1-R40 (2004).

\bibitem{barone}
A. Barone, G. Patern\'o, {\it Physics and applications of the
Josephson effect}, Wiley, New York (1982).

\bibitem{feld}
E. Feldtkeller, Phys. Stat. Sol. {\bf 27}, 161 (1968).

\bibitem{bar}
V.G. Bar'yakhtar, {\sl Dynamics of topological magnetic solitons}, 
Springer Tracts in Mod. Phys., Springer, Berlin (1994).

\bibitem{lamb}
G. L. Lamb, Rev. Mod. Phys. {\bf 43}, 99 (1971)

\bibitem{taylor}
J. R. Taylor, {\sl Optical Solitons}, Cambridge University Press, 
Cambridge (1992. 

\bibitem{cattuto}
C. Cattuto, F. Marchesoni, Phys. Rev. Lett. {\bf 79}, 5070 (1997).

\bibitem{manton}
N. Manton, and P. Sutcliffe, {\sl Topological solitons},
Cambridge University Press, Cambridge (2004).

\bibitem{lamb1}
G. L. Lamb, {\it Elements of Soliton Theory}, (Wiley, New York) (1980).

\bibitem{fm}
D. Finkelstein and C. W. Misner, Ann. Phys. {\bf 6}, 230 (1969).

\bibitem{skyrme}
T.H.R. Skyrme,  Proc. Roy. Soc. London A
{\bf 247}, 260 (1958).

\bibitem{zamo}
A.B. Zamolodchikov and A.B. Zamolodchikov Ann. Phys. {\bf 120} 253
(1979)

\bibitem{witten}
E. Witten, Nucl. Phys. {\bf B223}, 433 (1983).

\bibitem{witten1}
E. Witten, Nucl. Phys. {\bf B185}, 513 (1981).

\bibitem{Gegenberg:1997ns} J.~Gegenberg and G.~Kunstatter,
  %``Solitons and black holes,''
Phys.\ Lett.\ B {\bf 413}  274  (1997)[arXiv:hep-th/9707181].

\bibitem{Cadoni:1998ej}
 M.~Cadoni,
 %``2D extremal black holes as solitons,''
 Phys.\ Rev.\ D {\bf 58},  104001 (1998)
 [arXiv:hep-th/9803257].

 \bibitem{calogero} F. Calogero and A. Degasperis, {\it Spectral transform and
solitons}, North Holland (Amsterdam) 1982.

\bibitem{Zakh} V. Zakharov, {\it What is integrability ?}, Springer, Berlin 1991.

\bibitem{DNF} A. Dubrovin, S.P. Novikov and A. Fomenko, {\it Modern
geometry} (voll. I, II \& III), Springer (Berlin) 1984.

 
\bibitem{CDG}
M. Cadoni, R. De Leo and G. Gaeta,  q-bio.BM/0604014 (In press in 
Phys. Rev. E).

\bibitem{CDG1}
 M. Cadoni, R. De Leo and G. Gaeta, q-bio.BM/0604027 (In press in J. 
 Nonlinear Math. Phys.).
\bibitem{mpey}
M. Peyrard, B. Piette and W. J. Zakrzewski, Physica D {\bf 64},
355 (1993).
\bibitem{harikumar}
E. Harikumar, C. Nagaraja Kumar, and M. Sivakumar, Phys. Rev. D {\bf 
58}, 107703 (1998).
\bibitem{sarlet}
V. W. Sarlet, P.G.L. Leach and
F. Cantrijn,  Physica D {\bf 17}, 87 (1985).
 \bibitem{cooper}
F. Cooper, A. Khare, U. Sukhatme, Phys. Rep. {\bf 251}, 267 (1995).

 \bibitem{cooper1}
F. Cooper, A. Khare, U. Sukhatme, {\it Supersymmetry in  Quantum
Mechanics}, World Scientific, Singapore (2001).







\end{thebibliography}
\end{document}